\documentclass[letterpaper, 10 pt, conference]{ieeeconf}  

\usepackage{graphicx} 
\usepackage[T1]{fontenc}
\usepackage[latin9]{inputenc}
\usepackage{xcolor}
\usepackage{amsmath}
\usepackage{amssymb}
\usepackage{stackrel}

\DeclareMathOperator*{\esssup}{ess\,sup}

\usepackage{tikz}
\usepackage{algorithm,algorithmic}
\usepackage{booktabs}
\usepackage{arydshln}

\usepackage{float}

\usepackage[percent]{overpic}
\usepackage{rotating}
\usepackage{cite}

\IEEEoverridecommandlockouts                              

\makeatletter
\newtheorem{theorem}{Theorem}
\newtheorem{lemma}{Lemma}
\newtheorem{corollary}{Corollary}

\newtheorem{definition}{Definition}
\newtheorem{remark}{Remark}
\newtheorem{assumption}{Assumption}

\usepackage[T1]{fontenc}

\begin{document}

\title{Small-gain conditions for exponential incremental stability in feedback interconnections}
\author{Mohamed Yassine Arkhis$^{1}$, Denis Efimov$^{1}$\thanks{\noindent$^{1}$Inria, Univ. Lille, CNRS, UMR 9189 - CRIStAL, F-59000
Lille, France.\protect }}

\maketitle

\begin{abstract}
We prove that under a small-gain condition, an interconnection of two globally incrementally exponentially stable systems inherits this property on any compact connected forward invariant set. It is also demonstrated that the interconnection inherits a weaker version of incremental exponential stability globally. An example illustrating the theoretical findings is given. The example also shows that the uniform negativity of the Jacobian is not necessary for incremental exponential stability.
\end{abstract}

\section{\label{sec:Introduction} Introduction}
Contraction and incremental stability \cite{david}, \cite{demidovich1962dissipativity}, \cite{giaccagli2022incremental} are concepts that were introduced to study the behavior of trajectories of a possibly nonlinear system in relation to each other. Different works have addressed the problem of finding sufficient conditions for incremental stability of a system \cite{david}, \cite{demidovich1962dissipativity}, \cite{lohmiller1998contraction}, \cite{sontag2010contractive}. A recent paper \cite{forni2013differential} introduced a more detailed terminology: incremental stability (IS) which means that two arbitrary solutions stay close to each other (close with respect to the distance between their initial conditions) incremental asymptotic stability (IAS) which means IS with asymptotic convergence of solutions to each other, and incremental exponential stability (IES) which means IAS with an exponential rate of convergence. Although there are different kinds of incremental stability, most of the works that can be found in the literature, such as \cite{andrieu2016transverse}, \cite{kawano2024incremental}, \cite{sontag2010contractive}, focus on analyzing IES of nonlinear systems.\\There are two main approaches for studying the IES of a system. The first consists in proving that the Jacobian of the system satisfies some type of uniform negativity property \cite{lohmiller1998contraction}, \cite{sontag2010contractive}. The second one uses the Lyapunov approach. In the latter case, analyzing the incremental exponential stability of a nonlinear time-varying system:\begin{equation}\label{1}
    \dot{x}=f(t,x),
\end{equation}where \(x\in \mathbb{R}^n\) is the state, \(t\in\mathbb{R}\), and \(f:\mathbb{R}\times\mathbb{R}^n\to\mathbb{R}^n\) is a locally Lipschitz vector field, two types of Lyapunov functions can be used to formulate sufficient conditions of IES \cite{kawano2024incremental}. The first one is called an incremental Lyapunov candidate \cite{david}, which is a Lyapunov function \(V:\mathbb{R}^n\times \mathbb{R}^n\to\mathbb{R}\) for the system \eqref{1} and a copy of itself:\begin{equation}\label{2}
    \left\{\begin{aligned}
    \dot{x}_1=f(t,x_1),\\\dot{x}_2=f(t,x_2),\end{aligned}\right.
\end{equation}which directly targets the distance between two arbitrary solutions of \eqref{1}. The second one is called an exponential Finsler Lyapunov candidate \cite{forni2013differential} (the original name in \cite{forni2013differential} does not include the term 'exponential', which we added to emphasize that we are only treating the IES case), which is a Lyapunov function \(V:\mathbb{R}^n\times \mathbb{R}^n\to\mathbb{R}\) for the system \eqref{1} and its displacement dynamics:\begin{equation}\label{3}
    \dot{\delta x}= J_f(t,x) \delta x,
\end{equation}where \(\delta x \in \mathbb{R}^n,\ f\) is assumed to be a class \(\mathcal{C}^2\) vector field, and \(J_f\) is the Jacobian of \(f\) in \(x\), satisfying:\begin{equation}
    \begin{aligned}
        c_3 |\delta x|^2 &\leq V(x,\delta x) \leq c_4 |\delta x|^2,\\\dot{V}(x,\delta x)&\le -\alpha V(x,\delta x),
    \end{aligned}
\end{equation}\(\forall x,\delta x\in\mathbb{R}^n\), where \(\dot{V}\) is the derivative of \(V\) along solutions of \eqref{1}, \eqref{3}, and \(c_3,c_4,\alpha>0\) are positive constants (in the more generic cases of IS and IAS, \(\alpha\) is a class \(\mathcal{K}\) function \cite{forni2013differential}).\\ It was recently proved that in the case of autonomous systems, the existence of an incremental Lyapunov function for the system \eqref{1} is equivalent to the existence of an exponential Finsler Lyapunov function for \eqref{1}, \eqref{3} \cite{kawano2024incremental}. Although in the paper \cite{kawano2024incremental}, the autonomous systems are considered, it is easy to verify that the result can be developed for time-varying systems, and that a similar proof is valid.\\As with the classical notions of stability, generally, the higher is the dimension of the system, the more complex Lyapunov functions tend to be. Therefore, different decomposition tools are found to be very useful, then the results such as the small-gain theorem \cite{jiang1994small}, which allows one to show that under certain conditions, an interconnection inherits the stability properties of its subsystems, often, simplify the task of proving a stability property.\\ In this paper, we establish a small-gain theorem for an interconnection of two IES systems. We show that, under a small-gain condition, the interconnection of two IES systems is IES on any compact, connected, forward invariant set. We also relax the notion of incremental exponential stability, to obtain a global property. Finally, we provide a practical example showcasing the main result, with some practical simulations. The example shows that the uniform negativity property of the Jacobian is not necessary even in the case of the first order systems.

\textbf{Notation}
\begin{itemize}
    \item \(\mathbb{R}_+=\{x\in\mathbb{R}: x\ge 0\}\), where \(\mathbb{R}\) is the set of real numbers.
    \item \(|\cdot|\) is the Euclidean norm on \(\mathbb{R}^n\).
    \item For a Lebesgue measurable essentially bounded function \(u:\mathbb{R}\to \mathbb{R}^n\), we denote \(\|u\|_{\infty}^{\mathbb{A}}:=\esssup_{t\in\mathbb{A}}|u(t)|\) as the essential supremum of \(u\) over \(\mathbb{A}\subset \mathbb{R}\).
    \item \(\mathcal{K}\) denotes the set of continuous strictly increasing functions \(\gamma:\mathbb{R}_+\to\mathbb{R}_+\) with \(\gamma(0)=0\).
    \item \(\mathcal{K}_{\infty}\) denotes the set of radially unbounded functions \(\gamma\in\mathcal{K}\).
    \item \(\mathcal{KL}\) denotes the set of continuous functions \(\beta:\mathbb{R}_+\times \mathbb{R}_+\to \mathbb{R}_+\), such that for each \(s\in\mathbb{R}_+\), \(\beta(\cdot,s)\in\mathcal{K}\), and \(\beta(s,\cdot)\) is strictly decreasing to zero for all \(s>0\).
    \item A distance \(d:\mathcal{D}\times\mathcal{D}\to \mathbb{R}_+\) on the set \(\mathcal{D}\subset\mathbb{R}^n\) is a positive function that satisfies \(d(x,y)=0\) if and only if \(x=y\), and \(d(x,y)=d(y,x)\), and \(d(x,z)\le d(x,y)+d(y,z)\), for all \(x,y,z\in\mathcal{D}\).
\end{itemize}
\section{\label{sec:Preliminaries} Preliminaries}
This paper considers an interconnection of two nonlinear systems: \begin{equation}\label{System 1}
        \begin{pmatrix}
            \dot{x} \\ \dot{y}
        \end{pmatrix}=f(t,x,y)=\begin{pmatrix}
            f_1(t,x)+\rho_1 g_1(y)\\f_2(t,y)+\rho_2 g_2(x)
        \end{pmatrix}
\end{equation}where \(x\in\mathbb{R}^n\) and \(y\in\mathbb{R}^m\) are the state vectors, \(t\in\mathbb{R}\), \(\rho_1,\rho_2>0\) are interconnection gains that are supposed to be sufficiently small, \(f_1:\mathbb{R}\times \mathbb{R}^n\longrightarrow \mathbb{R}^n,g_2:\mathbb{R}^n \longrightarrow \mathbb{R}^m\) are \(\mathcal{C}^2\) in \(x\), and \(f_2:\mathbb{R}\times \mathbb{R}^m\longrightarrow \mathbb{R}^m,g_1:\mathbb{R}^m \longrightarrow \mathbb{R}^n\) are \(\mathcal{C}^2\) in \(y\). Furthermore, we assume that the system \eqref{System 1} is forward complete. Denote by \(f(t,z)=f(t,x,y)\) and \(J_f(t,z)=J_f(t,x,y)\) the Jacobian of \(f\) in the state \(z=(x,y)\in \mathbb{R}^{n+m}\).\\ For \(t_0\in\mathbb{R},z\in\mathbb{R}^{n+m}\), we denote \(\phi_{t_0}(\cdot,z)\) the solution of \eqref{System 1} with initial condition \(\phi_{t_0}(t_0,z)=z\).
The displacement dynamics along a solution \(\phi_{t_0}(\cdot,z)\) for \(z\in \mathbb{R}^{n+m}\) is given by:\begin{equation}\label{System 1 displacement}
    \dot{\delta z}(t)=J_f(t,\phi_{t_0}(t,z))\delta z(t).
\end{equation}
\subsection{Incremental exponential stability}
\begin{definition}\label{def: contraction}
    Let \(\mathcal{D}\subset \mathbb{R}^{n+m}\) be a forward invariant set for \eqref{System 1}. The system \eqref{System 1} is called:\begin{enumerate}
        \item \cite{forni2013differential} Incrementally exponentially stable (IES) on \(\mathcal{D}\) with respect to a distance \(d:\mathcal{D}\times \mathcal{D}\to \mathbb{R}_+\) if there exist \(K\ge 1, \lambda>0\) such that for all \(t_0\in\mathbb{R},z_1,z_2 \in \mathcal{D},t\ge t_0\): \[d(\phi_{t_0}(t,z_1),\phi_{t_0}(t,z_2))\le Ke^{-\lambda (t-t_0)}d(z_1,z_2).\]
        \item Weakly incrementally exponentially stable (WIES) on \(\mathcal{D}\) with respect to a distance \(d:\mathcal{D}\times\mathcal{D}\to \mathbb{R}_+\) if there exist \(\lambda>0,\gamma\in\mathcal{K}\) such that for all \(t_0\in\mathbb{R},z_1,z_2 \in \mathcal{D},t\ge t_0\): \begin{align*}d(\phi_{t_0}(t,z_1),\phi_{t_0}(t,z_2))\le \gamma(1+|z_1|+|z_2|)e^{-\lambda (t-t_0)}\\d(z_1,z_2).\end{align*}
    \end{enumerate}
    When \(\mathcal{D}=\mathbb{R}^{n+m}\), we say that system \eqref{System 1} is globally (W)IES. In the latter case the distance \(d\) can be replaced with a conventional norm in \(\mathbb{R}^{n+m}.\)
\end{definition}
\begin{definition}\cite{forni2013differential}\label{Finsler metric}
    For a connected set \(\mathcal{D}\in\mathbb{R}^{n+m}\), the function \(d:\mathcal{D}\times \mathcal{D}\to \mathbb{R}_+\) defined by: \[d(z_1,z_2):=\inf_{\gamma\in \Gamma(z_1,z_2)}\int_0^1 \left|\frac{\partial \gamma}{\partial s}(s)\right|ds,\]where \(\Gamma(z_1,z_2):=\{\gamma:[0,1]\to \mathcal{D}, \gamma\  \text{is}\  \mathcal{C}^1, \gamma(0)=z_1, \gamma(1)=z_2\}\) is a distance on \(\mathcal{D}\) called a Finsler distance induced by \(|\cdot|\) on \(\mathcal{D}\).
\end{definition}
\begin{remark}\label{remark 1}
    Notice that for any \(\mathcal{C}^1\) curve \(\gamma:[0,1]\to \mathbb{R}^{n+m}\) with \(\gamma(0)=z_1,\gamma(1)=z_2\): \[\int_0^1 \left|\frac{\partial \gamma}{\partial s}(s)\right|ds\ge \int_0^1 \left|\frac{\partial \theta}{\partial s}(s)\right|ds=|z_1-z_2|,\] where \(\theta(s):= sz_2+(1-s)z_1\) is the segment \([z_1,z_2]\). Thus, if \(\mathcal{D}\) is convex, then the Finsler distance induced by \(|\cdot|\) on \(\mathcal{D}\) is the Euclidean distance.
\end{remark}
\begin{theorem}\cite{forni2013differential}\label{theorem 1}
    Let \(\mathcal{D}\subset \mathbb{R}^{n+m}\) be a connected forward invariant set for \eqref{System 1}. If there exists an exponential Finsler Lyapunov function \(V:\mathbb{R}^{n+m}\times \mathbb{R}^{n+m}\to \mathbb{R}_+\), \(i.e.\), a \(\mathcal{C}^1\) function satisfying \(\exists \alpha, \underbar{c}, \bar{c}>0,\forall z\in \mathcal{D},\delta z \in \mathbb{R}^{n+m},t\in \mathbb{R}\): \begin{equation}\label{Finsler condition}
    \begin{aligned}
        \underbar{c}|\delta z|^2&\le V(z,\delta z) \le \bar{c}|\delta z|^2,\\\dot{V}(z,\delta z)&\le -\alpha V(z,\delta z),
    \end{aligned}\end{equation}where \(\dot{V}(z,\delta z):=\frac{\partial V(z,\delta z)}{\partial z}f(z,t)+\frac{\partial V(z,\delta z)}{\partial \delta z}J_f(z,t)\delta z\) is the derivative of \(V\) along solutions of \eqref{System 1}-\eqref{System 1 displacement}, and \(\frac{\partial V(z,\delta z)}{\partial z}\) is the gradient of \(V\) in \(z\), \(\frac{\partial V(z,\delta z)}{\partial \delta z}\) is the gradient of \(V\) in \(\delta z\), then the system \eqref{System 1} is IES on \(\mathcal{D}\) with respect to the Finsler distance induced by \(|\cdot|\) on \(\mathcal{D}\).
\end{theorem}
\begin{remark}\label{remark 2}
When \(\mathcal{D}\) is convex, condition \eqref{Finsler condition} implies IES on \(\mathcal{D}\) with respect to the Euclidean distance.
\end{remark}

\subsection{Input-to-state practical stability}
Consider a forward complete non-autonomous system:\begin{equation}\label{control system}
    \dot{x}=f(t,x,u),
\end{equation}where \(x\in \mathbb{R}^n,u\in\mathbb{R}^m\) and \(t\geq t_0\in\mathbb{R}\), \(f:\mathbb{R}\times\mathbb{R}^n\times\mathbb{R}^m\to\mathbb{R}^n\) is a continuous function.
\begin{definition}\cite{jiang1994small}\label{ISpS definition}
    The system \eqref{control system} is called input-to-state practically stable (ISpS), if there exist \(\gamma\in \mathcal{K},\beta\in \mathcal{KL},r>0\), such that \(\forall  t_0\in\mathbb{R},\  x(t_0)\in\mathbb{R}^n\), and any essentially bounded input \(u\), the respective solution \(x(\cdot)\) exists and satisfies \(\forall t\ge t_0\): \[|x(t)|\le \beta(|x(t_0)|,t-t_0)+\gamma(\|u\|_{\infty}^{[t_0,t]})+r.\]
\end{definition}
\begin{lemma}\label{lemma 1}
    A system \eqref{control system} is ISpS iff there exist \(q>0,\ \alpha_1,\alpha_2\in\mathcal{K}_{\infty},\ \chi,\alpha_3 \in \mathcal{K}\) and a smooth function \(V:\mathbb{R}\times \mathbb{R}^n\to\mathbb{R}_+\) such that \(\forall t\in\mathbb{R},\ x\in\mathbb{R}^n,\ u\in\mathbb{R}^m\):\begin{equation}\label{inequality 9}
        \begin{aligned}
            &\alpha_1(|x|)\le V(t,x) \le \alpha_2(|x|)\\&|x|\ge \max(\chi(|u|),q)\implies\dot{V}(t,x)\le -\alpha_3(|x|),
        \end{aligned}
    \end{equation}where \(\dot{V}(t,x)=\nabla V(t,x)(f(t,x,u),\ 1)^{\top}\), and \(\nabla V(t,x)\) is the gradient vector of \(V\). Furthermore, \(V\) is called an ISpS-Lyapunov function of the system \eqref{control system}.
\end{lemma}
\begin{proof}
    By definition, the system \eqref{control system} is ISpS, therefore there exist \(\gamma\in \mathcal{K},\beta\in \mathcal{KL},r>0\), such that \(\forall  t_0\in\mathbb{R}, x(t_0)\in\mathbb{R}^n\), and any piecewise continuous input \(u\), \(\forall t\ge t_0\): \begin{equation}\label{inequality 10}|x(t)|\le \beta(|x(t_0)|,t-t_0)+\gamma(\|u\|_{\infty}^{[t_0,t]})+r.\end{equation}
    We will use the main result of \cite{sontag2000lyapunov}. In \cite{sontag2000lyapunov} the autonomous systems are studied, thus, we consider the system \begin{equation}\label{system 10}
    \left\{\begin{aligned}
        \dot{x}(t)&=f(\Tilde{x}(t),x(t),u(t))\\\dot{\Tilde{x}}(t)&=1
    \end{aligned}\right.\end{equation}with output \(y(t)=x(t)\), and state \( [x^\top \tilde{x}]^\top\in\mathbb{R}^{n+1}\). The system \eqref{system 10} is autonomous and forward complete, and from Theorem 1.2 and Remark 4.2 of \cite{sontag2000lyapunov} the output \(x(t)\) satisfies inequality \eqref{inequality 10} iff there exists a Lyapunov function \(V:\mathbb{R}\times\mathbb{R}^n\to\mathbb{R}_+\) satisfying inequalities \eqref{inequality 9}.
\end{proof}
\begin{remark}
    Theorem 1.2 of \cite{sontag2000lyapunov} targets the input-to-output stability, but the result for input-to-output practical stability follows exactly the same reasoning.
\end{remark}
\section{Problem statement}
Our goal is to find sufficient conditions for the interconnection \eqref{System 1} to inherit the (W)IES property from its isolated subsystems.\\ 
We first assume that both subsystems have exponential Finsler Lyapunov candidates, which implies that both subsystems are IES.
\begin{assumption}
Both isolated systems \(\dot{x}=f_1(t,x),\ \dot{y}=f_2(t,y)\) admit Finlser Lyapunov functions for IES, \(i.e.\), \(\exists V_1:\mathbb{R}^n\times \mathbb{R}^n\longrightarrow \mathbb{R}_+,\ V_2:\mathbb{R}^m\times \mathbb{R}^m\longrightarrow \mathbb{R}_+\) satisfying \(\forall (x,y)\in \mathbb{R}^{n+m},(\delta x,\delta y)\in\mathbb{R}^{n+m}\): \begin{equation}\label{10}
\begin{aligned}
\underbar{c}_1|\delta x|^2 &\le V_1(x,\delta x) \le \bar{c}_1|\delta x|^2,\\\dot{V_1}(x,\delta x)&\le -\alpha_1 |\delta x|^2,
\end{aligned}
\end{equation}and: \begin{equation}\label{11}
    \begin{aligned}
        \underbar{c}_2|\delta y|^2 &\le V_2(y,\delta y) \le \bar{c}_2|\delta y|^2,\\
        \dot{V_2}(y,\delta y)&\le -\alpha_2 |\delta y|^2,\end{aligned}
\end{equation}where \( \underbar{c}_1,\underbar{c}_2,\bar{c}_1,\bar{c}_2,\alpha_1,\alpha_2>0\).
\end{assumption} 
It has been proven that conditions \eqref{10} and \eqref{11} are necessary and sufficient for IES in the case of autonomous systems \cite{kawano2024incremental}.
\begin{assumption}\label{assumption 2}
There exist continuous functions \( \gamma_1:\mathbb{R}^n\to \mathbb{R}_+,\ \gamma_2:\mathbb{R}^m\to \mathbb{R}_+,\ \zeta_1:\mathbb{R}^n\to \mathbb{R}_+,\ \zeta_2:\mathbb{R}^m\to \mathbb{R}_+\) such that \(\forall x,\delta x \in \mathbb{R}^n, y,\delta y \in \mathbb{R}^m\): \begin{equation}
        \begin{aligned}
            &\left|\frac{\partial V_1(x,\delta x)}{\partial x}\right|\le \gamma_1(x) |\delta x|^2&&,\ \left|\frac{\partial V_1(x,\delta x)}{\partial \delta x}\right|\le \zeta_1(x) |\delta x|,\\&\left|\frac{\partial V_2(y,\delta y)}{\partial y}\right|\le \gamma_2(y) |\delta y|^2&&,\ \left|\frac{\partial V_2(y,\delta y)}{\partial \delta y}\right|\le \zeta_2(y) |\delta y|.
        \end{aligned}
    \end{equation}
\end{assumption}
This assumption is a technical one. Among the exponential Finsler Lyapunov candidates, the simplest variants are quadratic forms: \(\delta x^{\top} M(x) \delta x\) with \(M(\cdot)\) being a \(\mathcal{C}^1\) symmetric matrix function \cite{lohmiller1998contraction}, and these clearly satisfy Assumption \ref{assumption 2}.
\section{Main results}
The first result shows that under assumptions 1-2, for small enough gains \(\rho_1\) and \(\rho_2\), the interconnection inherits IES on every compact connected forward invariant set. The second result shows that under assumptions 1-2, and an ultimate boundedness condition, the interconnection is globally WIES.
\begin{theorem}\label{Theorem 1}
    Under assumptions 1 and 2, for any compact, connected, forward invariant set \(\mathcal{D}\subset \mathbb{R}^{n+m}\) of \eqref{System 1}, there exist sufficiently small gains \(\rho_1, \rho_2>0\) such that the system \eqref{System 1} is IES on \(\mathcal{D}\) with respect to the Finsler distance induced by \(|\cdot|\) on \(\mathcal{D}\).\\When \(\mathcal{D}\) is convex, system \eqref{System 1} is IES on \(\mathcal{D}\) with respect to the Euclidean distance.
\end{theorem}

\begin{proof}
Let \(\mathcal{D}\subset\mathbb{R}^{n+m}\) be a compact, connected, forward invariant set for \eqref{System 1}, and let \(R>0\) be a positive number such that \(|\phi_{t_0}(t,z)|\le R\) for all \(t\ge t_0,z\in \mathcal{D}\). Recall that \(g_1, g_2\) are \(\mathcal{C}^2\), and \(\gamma_1,\gamma_2,\zeta_1,\zeta_2\) are continuous, therefore: \begin{equation*}
\begin{aligned}
&a_1=\sup_{|y|\le R}|g_1(y)|,a_2=\sup_{|x|\le R}|g_2(x)|,b_1=\sup_{|y|\le R}\left|\frac{\partial g_1(y)}{\partial y}\right|,\\&b_2=\sup_{|x|\le R}\left|\frac{\partial g_2(x)}{\partial x}\right|,\eta_1=\sup_{|x|\le R} |\gamma_1(x)|,\eta_2=\sup_{|y|\le R} |\gamma_2(y)|,\\&\vartheta_1=\sup_{|x|\le R} |\zeta_1(x)|,\vartheta_2=\sup_{|y|\le R} |\zeta_2(y)|.
\end{aligned}
\end{equation*} are real numbers.\\Consider a candidate exponential Finsler Lyapunov function: \[V(x,y,\delta x, \delta y):=V_1(x,\delta x)+V_2(y,\delta y).\]Denoting \(\underbar{c}=\min(\underbar{c}_1,\underbar{c}_2),\bar{c}=\max(\bar{c}_1,\bar{c}_2)\), using \eqref{10} and \eqref{11}, we have: \begin{equation}\label{V dot bounds}
    \underbar{c}|(\delta x, \delta y)|^2\le V(x,y,\delta x,\delta y)\le \bar{c}|(\delta x, \delta y)|^2.
\end{equation}The derivative along trajectories starting in \(\mathcal{D}\) of system \eqref{System 1}, \eqref{System 1 displacement} satisfies using \eqref{10} and \eqref{11} (the expressions for \(\dot{V}_1\) and \(\dot{V}_2\) are defined there): \begin{equation*}
            \begin{aligned}
             \dot{V}&\begin{aligned}[t]=&\dot{V_1}+\dot{V_2}+\rho_1 \frac{\partial V_1}{\partial x} g_1(y)+\rho_1 \frac{\partial V_1}{\partial \delta x}\frac{\partial g_1(y)}{\partial y} \delta y + \rho_2 \frac{\partial V_2}{\partial y}\\& \times g_2(x)+\rho_2 \frac{\partial V_2}{\partial \delta y}\frac{\partial g_2(x)}{\partial x} \delta x \end{aligned}\\& 
             \le \begin{aligned}[t] &-\alpha_1 |\delta x|^2-\alpha_2 |\delta y|^2+\rho_1\left(a_1 \eta_1 |\delta x|^2 + b_1 \frac{\vartheta_1^2}{2} |\delta x|^2+\right.\\&\left.\frac{b_1}{2} |\delta y|^2\right) + \rho_2 \left(a_2 \eta_2 |\delta y|^2 + b_2 \frac{\vartheta_2^2}{2} |\delta y|^2+ \frac{b_2}{2} |\delta x|^2\right)\end{aligned}\\&=\begin{aligned}[t]&\left(-\alpha_1+\rho_1\left(a_1\eta_1+\frac{b_1\vartheta_1^2}{2}\right)+\rho_2\frac{b_2}{2}\right)|\delta x|^2+\bigg(-\alpha_2\\&+\rho_2\left(a_2\eta_2+\frac{b_2\vartheta_2^2}{2}\right)+\left.\rho_1\frac{b_1}{2}\right)|\delta y|^2.\end{aligned}
             \end{aligned}
    \end{equation*}Therefore, for \(\rho_1,\rho_2>0\) sufficiently small satisfying: \begin{align}
        -\alpha_1+\rho_1\left(a_1\eta_1+\frac{b_1\vartheta_1^2}{2}\right)+\rho_2\frac{b_2}{2} \le -\alpha\\-\alpha_2+\rho_2\left(a_2\eta_2+\frac{b_2\vartheta_2^2}{2}\right)+\rho_1\frac{b_1}{2} \le -\alpha,
    \end{align}
    where \(\alpha>0\), we have for \((x,y)\in\mathcal{D},(\delta x,\delta y)\in \mathbb{R}^{n+m}\): \[\dot{V}\le -\alpha (|\delta x|^2+|\delta y|^2)=-\alpha |(\delta x,\delta y)|^2,\]We conclude from Theorem \ref{theorem 1} that 
    the system \eqref{System 1} is IES on \(\mathcal{D}\) with respect to the Finsler distance induced by \(|\cdot|\) on \(\mathcal{D}\).\\ From Remark \ref{remark 2} we conclude that if \(\mathcal{D}\) is convex, then the system \eqref{System 1} is IES on \(\mathcal{D}\) with respect to the Euclidean distance.
\end{proof}
\begin{remark}
    An example of a suitable pair \(\rho_1,\rho_2\) is:\begin{equation*}\begin{aligned}0<\rho_1\le \min\left( \frac{2\varepsilon_1}{2a_1\eta_1+b_1\vartheta_1^2},\frac{2\varepsilon_
    4}{b_1}\right),\\0<\rho_2\le \min\left( \frac{2\varepsilon_3}{2a_2\eta_2+b_2\vartheta_2^2},\frac{2\varepsilon_2}{b_2}\right),\end{aligned}\end{equation*}where \(\varepsilon_1,\varepsilon_2,\varepsilon_3,\varepsilon_4\) are any positive real numbers satisfying: \(\varepsilon_1+\varepsilon_2<\alpha_1-\alpha,\varepsilon_3+\varepsilon_4<\alpha_2-\alpha.\)
\end{remark}
\begin{corollary}\label{corollary 1}
If the $x$-subsystem in (5) is ISpS with a Lyapunov function (from Lemma \ref{lemma 1}) \(V_x\) with parameters \(r_x>0,\ \alpha_{1x},\alpha_{2x}\in\mathcal{K}_{\infty},\ \chi_x,\alpha_{3x} \in \mathcal{K}\) and the $y$-subsystem in (5) is ISpS with parameters \(r_y>0,\ \alpha_{1y},\alpha_{2y}\in\mathcal{K}_{\infty},\ \chi_y,\alpha_{3y} \in \mathcal{K}\), and a small gain condition holds:\begin{equation}
    \chi_x\circ\chi_y(r)\le r,\ \forall\ r>r_0>0,
\end{equation}then under assumptions 1-2, there exist a compact forward invariant set \(\mathcal{S}\subset \mathbb{R}^{n+m}\) and sufficiently small gains \(\rho_1, \rho_2\) such that the system \eqref{System 1} is IES on \(\mathcal{S}\) with respect to a Finsler distance induced by \(|\cdot|\) on \(\mathcal{S}\). 
\end{corollary}
\begin{proof}
    By introduced assumptions, \(V_x\) and \(V_y\) are ISpS-Lyapunov functions:\begin{equation}\label{inequality 20}
        \begin{aligned}
            &\alpha_{1x}(|x|)\le V_x(t,x) \le \alpha_{2x}(|x|)\\&|x|\ge \max(\chi_x(|y|),r_x)\implies\dot{V}_x(t,x)\le -\alpha_{3x}(|x|), 
        \end{aligned}
    \end{equation}and\begin{equation}\label{inequality 21}
        \begin{aligned}
            &\alpha_{1y}(|y|)\le V_y(t,y) \le \alpha_{2y}(|y|)\\&|y|\ge \max(\chi_y(|x|),r_y)\implies\dot{V}_y(t,y)\le -\alpha_{3y}(|y|).
        \end{aligned}
    \end{equation}From Theorem 3.1 of \cite{jiang1996lyapunov} we conclude that the interconnected system \eqref{System 1} is ISpS (the input is \(0\)). From Lemma \ref{lemma 1} there exist \(q>0,\ \alpha_1,\alpha_2\in\mathcal{K}_{\infty},\ \chi,\alpha_3 \in \mathcal{K}\) and a smooth function \(V:\mathbb{R}\times \mathbb{R}^{n+m}\to\mathbb{R}_+\) such that \(\forall t\in\mathbb{R},\ z=(x,y)\in\mathbb{R}^{n+m},\ u\in\mathbb{R}^m\):\begin{equation}
        \begin{aligned}
            &\alpha_1(|z|)\le V(t,z) \le \alpha_2(|z|)\\&|z|\ge q\implies\dot{V}(t,z)\le -\alpha_3(|z|),
        \end{aligned}
    \end{equation}
    then, taking a sufficiently large \(l>q\) there exists a forward invariant set \(\mathcal{S}\subset\{z=(x,y)\in\mathbb{R}^{n+m}: |z|\le l\}\) for \eqref{System 1}. Furthermore, \(\mathcal{S}\) is compact and connected. Thus, using Theorem \ref{Theorem 1}, there exist sufficiently small gains \(\rho_1, \rho_2>0\) such that the system \eqref{System 1} is IES on \(\mathcal{S}\) with respect to the Finsler distance induced by \(|\cdot|\) on \(\mathcal{S}\).
\end{proof}
\begin{remark}\label{remark 5}
    The small-gain condition is mainly assumed to show the existence of a compact forward invariant set (instead of small-gain restrictions for ISpS Lyapunov functions, similar conditions can be imposed on the time estimates). Therefore, if in an example, one can show directly the existence of a compact forward invariant set then the result of Corollary \ref{corollary 1} holds, as will be shown in the example section.
\end{remark}
\begin{theorem}
Suppose that the vector field \(f\) of system \eqref{System 1} is globally Lipschitz in \(z\), and that there exists a Lyapunov candidate \(W\) satisfying:\begin{equation}\label{18}
    \left\{\begin{aligned}
        &\alpha_1(|z|)\le W(t,z)\le \alpha_2(|z|),\ \forall z\in \mathbb{R}^{n+m},\\&\dot{W}(t,z)\le -\alpha_3(|z|),\ \forall |z|\ge \mu>0,
    \end{aligned}\right.
\end{equation}where \(\alpha_1,\alpha_2,\alpha_3\in \mathcal{K}_{\infty}\), and \(\dot{W}\) is the derivative of \(W\) along the trajectories of system \eqref{System 1}.
Then, under assumptions 1-2, there exist sufficiently small gains \(\rho_1,\rho_2\) such that the system \eqref{System 1} is globally WIES.
\end{theorem}
\begin{proof}
The existence of \(W\) proves that the set \(\mathcal{S}=\{z\in\mathbb{R}^{n+m}: |z|\le R\}\), for \(R>\alpha_1^{-1}\circ\alpha_2(\mu)\), is forward invariant for the system \eqref{System 1}. \(\mathcal{S}\) is also compact and convex, therefore, from Theorem \ref{Theorem 1}, there exist sufficiently small gains \(\rho_1, \rho_2\) such that the associated system \eqref{System 1} is IES on \(\mathcal{S}\) with respect to the Euclidean distance, \(i.e.\), there exist \(K\ge 1\) and \(\alpha>0\) such that for all \(t_0\in\mathbb{R},z_1,z_2 \in \mathcal{S}\): \begin{equation}\label{22}|\phi_{t_0}(t,z_1)-\phi_{t_0}(t,z_2)|\le Ke^{-\alpha (t-t_0)}|z_1-z_2|,\ \forall t\ge t_0.\end{equation}Let \(z_1,z_2 \in \mathbb{R}^{n+m}\). From Theorem 4.18 of \cite{khalil2002control}, the condition \eqref{18} implies that there exists \(T>0\) dependent only on \(R,z_1,z_2\) such that \(\forall t_0\in \mathbb{R}\) and all \(t\ge t_0+T\):\begin{equation}
    |\phi_{t_0}(t,z_1)|\le R, \text{ and, }|\phi_{t_0}(t,z_2)|\le R,
\end{equation}therefore \(\forall t\ge t_0+T\):\begin{equation}\label{23}
\begin{aligned}
        |\phi_{t_0}(t,z_1)-\phi_{t_0}(t,z_2)|\le Ke^{-\alpha (t-T-t_0)}|\phi_{t_0}(t_0+T,z_1)\\-\phi_{t_0}(t_0+T,z_2)|.\end{aligned}
    \end{equation}Recalling that \(f\) is globally Lipschitz in \(z\), there exists \(\lambda>0\) such that \(\forall t_0\le t\le t_0+T\): \[|\phi_{t_0}(t,z_1)-\phi_{t_0}(t,z_2)|\le e^{\lambda T}|z_1-z_2|.\]Combining this with inequality \eqref{23}, we obtain for all \(t_0 \in \mathbb{R}, t \ge t_0\):\begin{equation}
        |\phi_{t_0}(t,z_1)-\phi_{t_0}(t,z_2)|\le Ke^{(\alpha+\lambda) T}e^{-\alpha (t-t_0)}|z_1-z_2|.
    \end{equation}From the continuous dependence of the solutions on initial conditions, we know that \(T\) depends continuously on initial conditions \(z_1,z_2\), therefore, there exists \(\gamma\in \mathcal{K}\), such that for any \(z_1,z_2\in\mathbb{R}^{n+m}\), the associated time \(T\) satisfies: \(T\le \gamma(1+|z_1|+|z_2|)\). We conclude that system \eqref{System 1} is globally WIES with respect to the Euclidean distance.
\end{proof}
\begin{remark}
    Notice that although the WIES is a weaker version of IES, it is relevant, since it shows that the trajectories are converging globally towards each other with an exponential speed that is independent on initial conditions (the power of the exponential in the upper bound does not depend on initial conditions.)
\end{remark}
\section{Example}
    Consider the FitzHugh-Nagumo model of an excitable system, which is a two-dimensional simplification of the Hodgkin-Huxley model of spike generation in squid giant axons: \begin{equation}\label{Fritz model}
        \left\{\begin{aligned}
            \dot{x}&=x-\frac{x^3}{3}+c-\rho_1 y\\ \varepsilon \dot{y}&=-by+\rho_2 x,
        \end{aligned}\right.
    \end{equation}where \(b,\rho_1,\rho_2,\varepsilon>0,c=r^3-r,r>2\).\\First we prove that the two isolated systems: \begin{equation}\label{Subsystem 1}
        \dot{x}=x-\frac{x^3}{3}+c,
    \end{equation} \begin{equation}\label{Subsystem 2}
        \varepsilon\dot{y}=-by,
    \end{equation} have respective exponential Finsler Lyapunov functions.\\Noticing that the Jacobian of the second system \(J_y(y)=-\frac{b}{\varepsilon}\) is uniformly negative definite, we conclude that the system \eqref{Subsystem 2} is IES on \(\mathbb{R}\) due to Theorem 1 in \cite{sontag2010contractive}, and the quadratic exponential Finsler Lyapunov function \(V_2(y,\delta y):=\frac{1}{2}\delta y^2\) can be used. The Jacobian of the first system \(J_x(x)=1-x^2\) is not uniformly negative definite, therefore, we cannot conclude directly that the system is IES, and we need to construct an exponential Finsler Lyapunov function. Define:\begin{equation}
        f_c(x)=\left\{\begin{aligned}
           &e^{\displaystyle \int_{\sqrt{\frac{2+\alpha}{2}}}^x\frac{2s^2-2-\alpha}{s-\frac{s^3}{3}+c}ds} &&,\ \text{for } |x|<\sqrt{\frac{2+\alpha}{2}},\\&1  &&,\ \text{for }x\ge \sqrt{\frac{2+\alpha}{2}},\\&e^{\mu}  &&,\ \text{for }x\le -\sqrt{\frac{2+\alpha}{2}},
        \end{aligned}\right.
    \end{equation}where \(0<\alpha < 2r^2-2\) and \(\mu=-\displaystyle \int_{-\sqrt{\frac{2+\alpha}{2}}}^{\sqrt{\frac{2+\alpha}{2}}}\frac{2s^2-2-\alpha}{s-\frac{s^3}{3}+c}ds\). Note that \(f_c\) is \(\mathcal{C}^1\) (since it is \(\mathcal{C}^1\) on \(\mathbb{R}\backslash\{-\sqrt{\frac{2+\alpha}{2}},\sqrt{\frac{2+\alpha}{2}}\}\) and \(f_c'(x)\to 0\) when \(x\to \pm \sqrt{\frac{2+\alpha}{2}}\)), and for \(x\in \mathbb{R}\):\begin{equation}\label{derivative of fc}
        f_c'(x)=\left\{\begin{aligned}
            &\frac{2x^2-2-\alpha}{x-\frac{x^3}{3}+c}f_c(x)&&,\text{ for }|x|<\sqrt{\frac{2+\alpha}{2}}\\&0&&,\text{ for }|x|\ge \sqrt{\frac{2+\alpha}{2}}
        \end{aligned}\right.
    \end{equation}
Notice that \(x-\frac{x^3}{3}+c>0 \iff x<r\), which implies that \(f_c\) is decreasing. We have \(\forall x\in \mathbb{R}^n\):\begin{equation}\label{f bounds}
    1\le f_c(x) \le e^{\mu},\ 
    -\eta\le f_c'(x)\le 0,
\end{equation}where \(\eta=-\min_{|x|\le \sqrt{\frac{2+\alpha}{2}}}f_c'(x)>0\).\\Consider an exponential Finsler Lyapunov candidate: \(V_1(x,\delta x)=f_c(x)\delta x^2\).\\
The derivative of \(V_1\) along the trajectories of the isolated subsystem \begin{equation}\label{dis 1}
    \left\{\begin{aligned}
        \dot{x}&=x-\frac{x^3}{3}+c\\\dot{\delta x}&=(1-x^2)\delta x
    \end{aligned}\right.
\end{equation}
is given by: \begin{align*}
    \dot{V_1}(x,\delta x)&=f_c'(x)(x-\frac{x^3}{3}+c)\delta x^2+f_c(x)\delta x^2 (2-2x^2)\\&=V_1(x,\delta x) (\frac{f_c'(x)}{f_c(x)}(x-x^3+c)+2-2x^2)
\end{align*}
For \(|x|\ge \sqrt{\frac{2+\alpha}{2}}\), \(f_c'(x)=0\) and \(2-2x^2\le -\alpha\).\\For \(|x|< \sqrt{\frac{2+\alpha}{2}}\), \(\frac{f_c'(x)}{f_c(x)}=\frac{2x^2-2-\alpha}{x-\frac{x^3}{3}+c}\), which implies that \(\frac{f_c'(x)}{f_c(x)}(x-\frac{x^3}{3}+c)+2-2x^2=-\alpha\). As a result \(\forall x,\delta x \in \mathbb{R}^n\): \begin{equation}\label{contracting 1 inequality}
    \dot{V_1}(x,\delta x)\le -\alpha V_1(x,\delta x),
\end{equation} and the system \eqref{Subsystem 1} is IES on \(\mathbb{R}\).\\Next, recalling Remark \ref{remark 5} we prove the existence of a  compact forward invariant set for the system \eqref{Fritz model} for any \(\rho_1=\rho_2>0\). For this purpose, consider a candidate Lyapunov function: \(W(x,y)=\frac{1}{2}(x^2+\varepsilon y^2)\). The derivative of \(W\) along trajectories of system \eqref{Fritz model} is given by: \begin{equation}\label{State bounds}
\begin{aligned}
    \dot{W}(x,y)&=x(x-\frac{x^3}{3}+c-\rho_1 y)+y(-by+\rho_2 x)\\&\le \frac{3}{2}x^2-\frac{x^4}{3}+\frac{c^2}{2}-by^2\\&\le -\frac{1}{8}x^2-by^2+2+\frac{c^2}{2}\\&\le -2\min(\frac{1}{8},\frac{b}{\varepsilon})W(x,y)+2+\frac{c^2}{2}.
\end{aligned}
\end{equation}Clearly, for large \(|x|\) and \(|y|\), the derivative of \(W\) is negative, therefore, for sufficiently large \(r>0\), there exist a compact connected forward invariant set for the system \eqref{Fritz model} \(\mathcal{S}\subset\{z=(x,y)\in\mathbb{R}^{n+m}: |x|\le r,\ |y|\le r\}\). As a result of Corollary \ref{corollary 1}, we conclude that for sufficiently small gains \(\rho_1,\ \rho_2\) the system \eqref{Fritz model} is IES on \(\mathcal{S}\).\\Notice that by applying the steps of the proof, we can obtain precise conclusion of when the system \eqref{Fritz model} is IES (globally), because we have complete knowledge of the upper-bounds. In fact, for \(\rho_1=\rho_2=1\), we can obtain conditions on the parameters of the system to be IES. Considering the same Lyapunov candidate as in the proof of Theorem \ref{Theorem 1}: \[V(x,y,\delta x, \delta y)=V_1(x,\delta x)+V_2(y,\delta y),\] we have:\begin{equation}
    \dot{V}\le -\alpha |\delta x|^2 -\frac{b}{\varepsilon} |\delta y|^2 + \eta |y| |\delta x|^2 +\left(2 e^{\mu}+\frac{1}{\varepsilon}\right) |\delta x||\delta y|
\end{equation}
From \eqref{State bounds} and the definition of \(W\), in the case of \(b>\varepsilon\), we obtain, from the comparison lemma \cite{khalil2002control}, for any \(M>0\) and \(T>0\) and any solution \((x(t),y(t))\) such that \(\frac{1}{2}(x(0)^2+\varepsilon y(0)^2)\in \mathbb{R}^{n+m}\), we have \(\forall t\ge T\):\begin{equation}\label{Fritz bounds}
\varepsilon y(t)^2\le\frac{\varepsilon}{b} (1+\frac{c^2}{4})+M.\end{equation}From \eqref{f bounds}, \eqref{Fritz bounds} and \eqref{contracting 1 inequality} we obtain: \begin{equation}\label{30}
\begin{aligned}
    \dot{V}\le\left(-\alpha+\eta \sqrt{\frac{1}{b}(1+\frac{c^2}{4})+\frac{M}{\varepsilon}}+\frac{e^{\mu}}{\varepsilon}+\frac{1}{2}\right)|\delta x|^2+\left(-\frac{b}{\varepsilon}\right.\\\left.+\varepsilon e^{\mu}+\frac{1}{2}\right)|\delta y|^2,\end{aligned}
\end{equation}notice that by choosing \(\varepsilon\) and \(b\) large enough with \(b\) larger than \(\varepsilon\) to compensate \(\varepsilon e^{\mu}\) the system is IES, as will be shown in the simulations.
\begin{remark}
    The example does not only showcase Theroem \ref{Theorem 1}, but it also demonstrates that the uniform negativity of the Jacobian is not necessary to induce contraction, by proving that system \eqref{Subsystem 1} is IES on \(\mathbb{R}\).
\end{remark}
\section{Numerical simulation}
\begin{figure}
\includegraphics[width=0.53\textwidth]{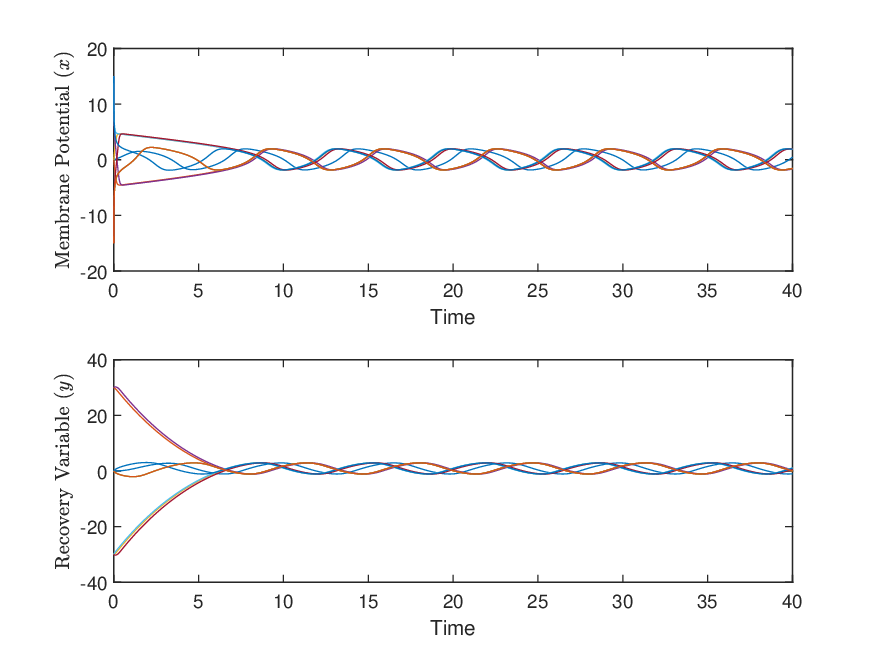}
\caption{\(c=1,b=0.1,\varepsilon=1,\rho_1=\rho_2=1\).}
\label{fig1}
\end{figure}

\begin{figure}
\includegraphics[width=0.53\textwidth]{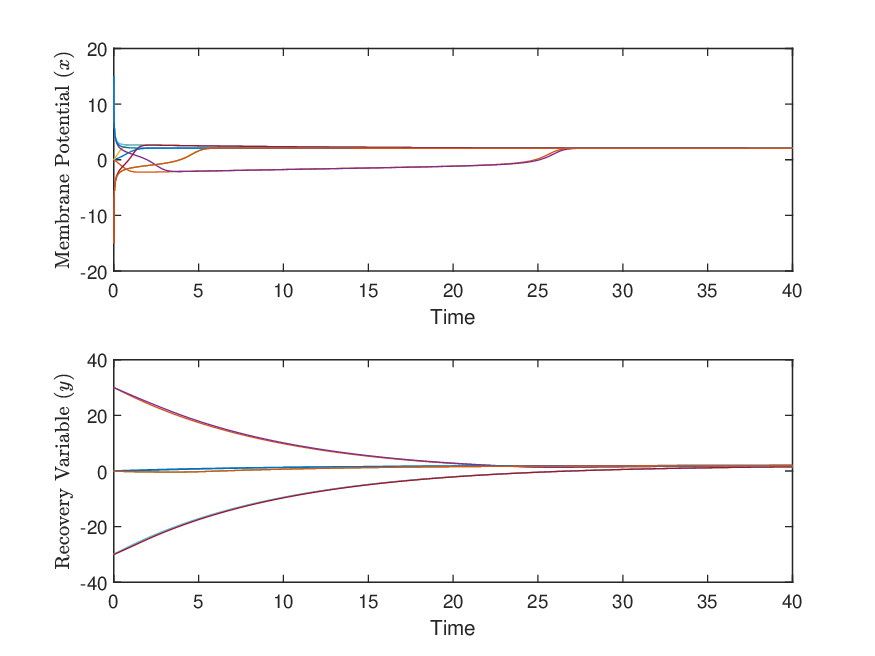}
\caption{\(c=1,b=0.1,\varepsilon=1,\rho_1=\rho_2=0.1\).}\label{fig2}
\end{figure}

\begin{figure}
\includegraphics[width=0.53\textwidth]{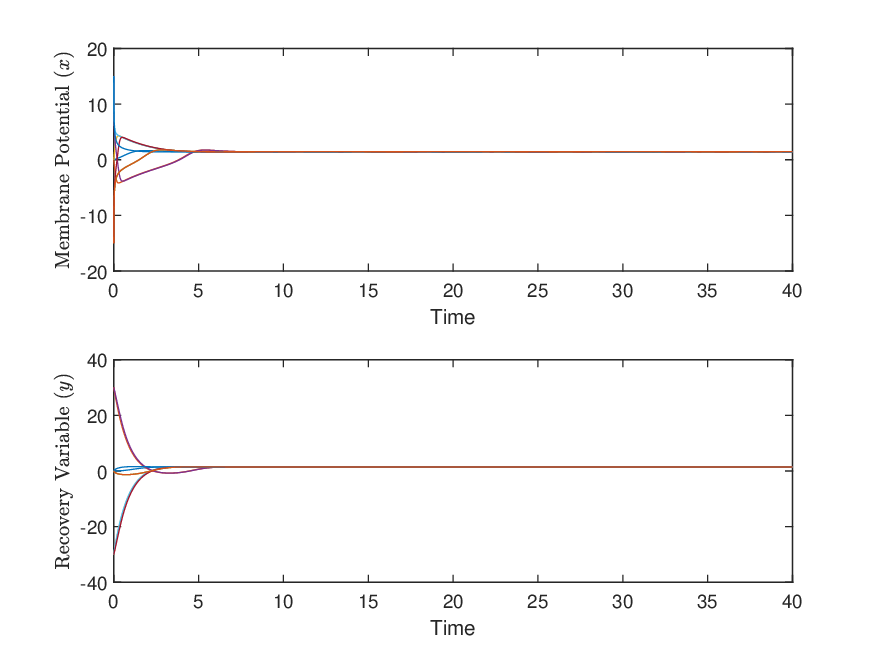}
\caption{\(c=1,b=1,\varepsilon=0.9,\rho_1=\rho_2=1\).}\label{fig3}
\end{figure}

First, Figure \ref{fig1} shows that even when the two separated subsystems are globally IES, under arbitrary big \(\rho_1,\rho_2\), the interconnection is not IES (the values of used parameters are indicated in the caption). Second, Figure \ref{fig2} illustrates the result of Theorem \ref{Theorem 1}, that for sufficiently small \(\rho_1\) and \(\rho_2\), the interconnection inherits the IES property. Finally, Figure \ref{fig3} demonstrates what we remarked from inequality \eqref{30}, for sufficiently large \(\varepsilon\) and larger \(b\), the interconnection is IES.

\section{Conclusion}
In this work we proved that under a small-gain condition, an interconnection of two (W)IES subsystems, inherits IES on every compact, connected, forward invariant set, and global WIES, which shows that all trajectories approach each other with the same exponential speed. We also highlighted through an example that uniform negativity of the Jacobian is not necessary for incremental stability. As a possible direction for future research, establishing a necessary condition or a converse theorem - one that shows that an IES system, under certain conditions, has an exponential Finsler Lyapunov function - can be considered.
\bibliographystyle{plain}
\bibliography{main}

\end{document}